\documentclass[12pt]{article}

\usepackage{subfig}
\usepackage{tabularx}
\usepackage{enumitem}
\usepackage{pifont}     
\usepackage{subcaption}
\usepackage{makecell}
\usepackage{longtable}
\usepackage{placeins}
\usepackage{float}
\usepackage{cite}
\usepackage{amsmath,amssymb,amsfonts}
\usepackage[font=small]{caption}
\usepackage{algorithmic}
\usepackage{graphicx}
\usepackage{textcomp}
\usepackage{xcolor}
\usepackage{tcolorbox}
\usepackage{url}
\usepackage[colorlinks=false, pdfborder={0 0 0}]{hyperref}

\newcommand{\cmark}{\ding{51}}
\newcommand{\xmark}{\ding{55}}
\begin{document}

\date{}

\title{Content and Quality Analysis of mHealth Apps for Feeding Children with Autism Spectrum Disorder}

\author{
Christopher Cofie Kuzagbe\textsuperscript{1}, 
Fabrice Mukarage\textsuperscript{1}, 
Skye Nandi Adams\textsuperscript{2},\\
N'guessan Yves-Roland Douha\textsuperscript{1},
Edith Talina Luhanga\textsuperscript{1}\thanks{Corresponding author: Edith Talina Luhanga (eluhanga@andrew.cmu.edu)} \\
\\
\textsuperscript{1}Carnegie Mellon University Africa, Kigali, Rwanda\\
\textsuperscript{2}University of the Witwatersrand, Johannesburg, South Africa\\
}

\maketitle


\begin{abstract}
\textit{Background:} Approximately 1 in 100 children worldwide are diagnosed with Autism Spectrum Disorder (ASD), and 46\% to 89\% experience significant feeding difficulties. Mobile health applications (mHealth apps) have emerged as a potential tool for scalable support. However, their quality and relevance in managing ASD-related feeding challenges remain unclear.

\textit{Objective:} To identify and evaluate the quality of mHealth apps available in the Africa region addressing feeding difficulties in children with ASD.

\textit{Methods:} A systematic search was conducted on the Apple App Store and Google Play Store between September and October 2024. Applications were included if they were free, in English, updated within the past year, explicitly focused on feeding in children with autism, available in the Africa region, and had more than 100 downloads. Eligible apps were assessed using the Behavior Change Wheel (BCW) framework and rated with the Mobile App Rating Scale (MARS) across four domains: engagement, functionality, aesthetics, and information quality.

\textit{Results:} Of the 326 applications identified, only two iOS apps met all inclusion criteria. EduKitchen–Toddlers Food Games featured child-centered interactive games and sensory-friendly visuals, while Autism Food Coach 2 provided structured caregiver tools, visual meal plans, and progress tracking. Both apps aligned with multiple BCW intervention functions, including education, training, and enablement. MARS scores of 3.7 and 3.9 indicated acceptable to good usability and content quality. The apps' shortcomings included limited customization for diverse user needs and the absence of documented clinical validation.

\textit{Conclusion:} There is a critical shortage of mHealth apps for feeding difficulties in children with ASD that are both evidence-based and of high quality. Future development must integrate robust clinical validation and comprehensive, caregiver-centered support features to address this significant gap.

\vspace{1em}
\end{abstract}

\section{Introduction} 
Autism Spectrum Disorder (ASD) is a complex neurodevelopmental condition characterized by challenges in social interaction, communication, behavior regulation, and sensory processing. The global prevalence of ASD has increased significantly in recent decades; the U.S. Centers for Disease Control and Prevention (CDC) reports a 178\% increase in autism prevalence since 2000 \cite{cdc_autism_2023}, and the World Health Organization (WHO) estimates that approximately 1 in 100 children is diagnosed with ASD \cite{who_autism_2023}.

Among the various comorbidities observed in children with ASD, feeding difficulties are especially common and under-addressed. Studies report prevalence rates ranging from 46\% to 89\% \cite{nadon_feeding_2013}, with affected children exhibiting food selectivity, aversions to textures and flavors, rapid eating, and disruptive mealtime behaviors \cite{marshall_features_2014, leader_feeding_2020}. These challenges are linked to nutritional deficiencies, impaired growth, gastrointestinal disorders, and metabolic imbalances \cite{xiao_correlation_2016, marinov_micronutrient_2024}. Children with ASD are estimated to be five times more likely to experience feeding problems than their neurotypical peers \cite{sharp_feeding_2013}. This situation places a significant burden on caregivers and parents, who often serve as the primary facilitators of feeding interventions and manage complex routines while navigating behavioral and communication challenges.

Parent-mediated interventions have become increasingly central in ASD care, with research highlighting their effectiveness in improving child outcomes and enhancing parent-child interactions \cite{Rojas-Torres2020, Beaudoin2014}. The emergence of mobile health applications (mHealth apps) presents new opportunities to deliver such interventions, particularly in resource-constrained or remote settings. These mHealth apps provide potentially accessible and cost-effective support for parents of children with ASD through psychoeducation, skill training, and behavior management guidance \cite{bharat_mhealth_2022, Dunn2017}. 

Several studies have reviewed mobile applications that address autism broadly\cite{aziz_study_2019}, including those focused on early intervention \cite{bharat_mhealth_2022}, communication support \cite{law_the_2018}, and parental empowerment \cite{bonnot_mobile_2021}. Some digital interventions also focus on the sensory or behavioral aspects of mealtime challenges; however, their quality and effectiveness remain inconsistent, often due to limited regulatory oversight and a lack of clinical validation \cite{larco_review_2018}. Furthermore, digital tools that do not align with parents’ expectations, parenting styles, or cultural contexts can exacerbate the stress experienced by caregivers of children with ASD \cite{Bradshaw2022, Houser2014}. Consequently, a more focused evaluation is needed. While existing reviews offer valuable insights, they are broad in scope and do not specifically examine mobile applications designed to address feeding difficulties in children with ASD. To address this gap, the present study systematically evaluates the content and quality of mobile applications designed to support feeding interventions for children with ASD.

This study aims to achieve the following objectives:
\begin{enumerate}
    \item Conduct a systematic review of mHealth apps specifically designed to address feeding challenges in children with ASD.
    \item Analyze the behavior change potential of the included apps using the Behavior Change Wheel (BCW) \cite{michie_behaviour_2011}, which is a framework to characterize behavior change interventions.
    \item Assess the quality of these mHealth apps using the Mobile Application Rating Scale (MARS) \cite{stoyanov2015mobile}, which evaluates engagement, aesthetics, functionality, and information quality.
    \item Identify the strengths and limitations of existing apps in supporting feeding-related interventions.
    \item Provide evidence-based, actionable recommendations to guide the development of future mHealth apps that more effectively assist caregivers in managing ASD-related feeding difficulties.
\end{enumerate}

\section{Methods}
\label{methods}
\subsection{Search Strategy}
We conducted a systematic search of the Apple App Store (iOS) and Google Play Store (Android) to identify mHealth apps related to feeding, food, meal planning, or nutrition for children with autism. The search was performed between September and October 2024. The search terms used included: ``Child autism food,'' ``Kids autism food,'' ``Toddler autism food,'' ``Child autism nutrition,'' ``Kids autism nutrition,'' ``Toddler autism nutrition,'' ``Child autism meal prep planning,'' ``Kids autism meal prep planning,'' ``Toddler autism meal prep planning,'' ``Child autism feeding,'' ``Kids autism feeding,'' ``Toddler autism feeding,'' ``Child autism mealtime,'' ``Kids autism mealtime,'' and ``Toddler autism mealtime.'' These terms were entered directly into the App Store and Google Play search interfaces. To reflect regional availability, the devices' region settings were configured to Africa during the search. The searches were conducted using an iPhone 12 Pro Max (iOS) and a Google Pixel 3a XL (Android) to simulate a typical end-user experience on each platform. All returned results were saved in an Excel file for deduplication and screening. 

\subsection{Application Screening}
After removing duplicates, two members of the research team independently reviewed the applications' information pages (screenshots and textual description) and screened them against the inclusion and exclusion criteria. Disagreements were resolved by a third member of the research team. 

\subsubsection{Inclusion Criteria}
Applications were included if (1) the content was in English; (2) the title or description explicitly addressed feeding in children with autism; (3) the app was free to download (either fully free or offering in-app purchases); (4) the app has more than 100 downloads; (5) the app had been updated within the past year (to exclude defunct or outdated apps).

We limited inclusion to English-only applications, as it is the most widely spoken language worldwide when combining native and non-native speakers. This choice is particularly relevant in Africa, where English is commonly used as an official or working language in many countries, making English-language apps easier for researchers and caregivers in the region \cite{ethnologue_english_2024}. We also aimed to include only applications that are easily discoverable by users (ranked higher in the results). Search results are ranked using various factors, including user engagement, and the number of downloads is one of the metrics used to determine engagement. We only included applications with more than 100 downloads, as initial searches showed that these were more likely to appear higher in the search results. This approach had the added advantage of reducing the risk of including harmful, low-quality, or abandoned applications \cite{huckvale2015smartphone}.

\subsubsection{Exclusion Criteria}
Applications were only excluded if they could not be installed on test devices (iPhone 12 Pro Max for iOS and Google Pixel 3a XL for Android), either due to geographical restrictions or other installation challenges. The former represents a practical limitation of our methodology, as it may have led to the omission of potentially suitable apps developed for other markets like America, Asia, or Europe. 

\subsection{Content Analysis}
Applications that met the inclusion criteria were downloaded and independently analyzed by two reviewers, one focusing on the Apple App Store and the other on the Google Play Store. To enhance reliability, reviewers cross-validated each other's assessments by switching platforms after the initial evaluations. For each application, all freely accessible screens were navigated, and informational content and interactive features were documented in an Excel table.

Deductive coding was then conducted using the Behavior Change Wheel (BCW) \cite{michie_behaviour_2011}, which categorizes behavioral determinants into deficits in capacity, opportunity, or motivation, and systematically maps these to nine intervention functions: \textit{education}, \textit{persuasion}, \textit{incentivization}, \textit{coercion}, \textit{training}, \textit{restriction}, \textit{environmental restructuring}, \textit{modeling}, and \textit{enablement}. In a preparatory session, both reviewers reviewed the BCW framework and developed a shared description of intervention functions. They also created examples of app features to ensure consistent coding. Discrepancies were discussed until consensus was reached; when consensus was not possible, a third reviewer was consulted. A deductive approach was selected to assess the extent to which app content aligned with established behavior change theory, rather than to generate novel thematic categories.

\subsection{Application Quality Assessment}
The quality of the applications was assessed using the Mobile App Rating Scale (MARS) \cite{stoyanov2015mobile}, a validated tool for the evaluation of mHealth apps. MARS evaluates applications on four dimensions: engagement (user interest and interaction), functionality (performance and usability), aesthetics (visual design), and quality of information (accuracy, relevance and credibility of content). Each dimension is rated on a 5-point Likert scale. Two reviewers independently rated each application. Both reviewers have a background in mobile application design and development and had worked on mHealth projects over the past year. The reviewers also completed the official MARS training to ensure consistent interpretation of the scale.

Following the independent rating phase, the two reviewers discussed their evaluations to reach a consensus score for each MARS item. These final consensus scores were used in the analysis.

\begin{figure}[t]
    \centering
\includegraphics[width=\textwidth]{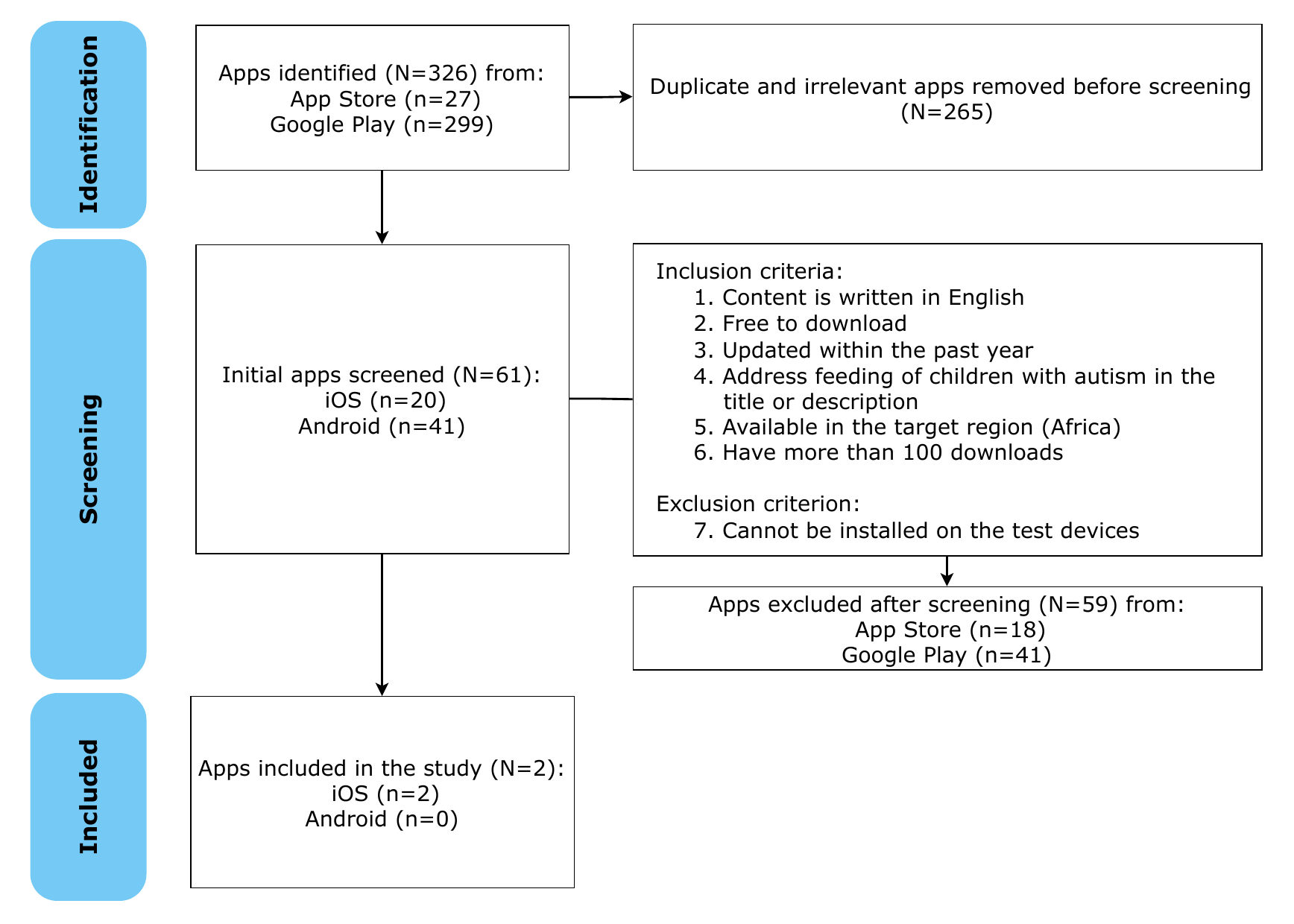}  
    \caption{PRISMA flow diagram of mHealth app selection for the systematic review.}
    \label{Fig:PRISMA}
\end{figure}

\section{Results}
\label{results} 

\subsection{Application Search and Selection}

The initial search yielded 326 applications (27 iOS apps from the Apple App Store and 299 Android apps from Google Play). After removing duplicates, the remaining applications were screened against the inclusion criteria. Most applications were excluded during this process, leaving only two iOS applications (\textit{EduKitchen–Toddlers Food Games} and \textit{Autism Food Coach 2}) that explicitly addressed feeding challenges in children with autism and met all inclusion criteria. No Android applications met the criteria. Figure \ref{Fig:PRISMA} presents the application selection process following the Preferred Reporting Items for Systematic Reviews and Meta-Analyses (PRISMA) guidelines \cite{liberati_prisma_2009}.

Most of the excluded applications focused on broader autism support, such as augmentative and alternative communication (AAC), social or behavioral skills, language development, or general early learning without specifically addressing feeding. A complete list of the screened applications, including their platform and primary focus, is provided in \hyperref[appendixScreenedApps]{Appendix}.


\subsection{Characteristics of Selected Applications}

Two applications met all inclusion criteria. \textit{EduKitchen–Toddlers Food Games} targets young children directly while \textit{Autism Food Coach 2} focuses on caregivers supporting feeding routines. Table~\ref{tab_coreFeatures} summarizes their core features and access models.

\begin{table}[H]
    \centering
    \caption{Core features and access models of the selected apps.}
    \renewcommand{\arraystretch}{1}
    \begin{tabular}{|p{2.5cm}|p{2.8cm}|p{7cm}|}
        \hline
        \textbf{App name} & \textbf{Access model} & \textbf{Core features} \\
        \hline
        \textit{EduKitchen – Toddlers Food Games} & One-time purchase &
        \begin{minipage}[t]{\linewidth}
        \begin{itemize}[leftmargin=*, nosep, topsep=0pt, itemsep=0pt, partopsep=0pt, after=\vspace{4pt}]
            \item Gamified food-related activities (e.g., sorting, matching, counting)
            \item Reinforcement through animations and in-app rewards
            \item Customizable settings for sensory preferences and learning levels
            \item  Engaging, child-friendly interface with cartoon visuals
        \end{itemize}
        \end{minipage} \\
        \hline
        \textit{Autism Food Coach 2} & Subscription (£49.99/month) &
        \begin{minipage}[t]{\linewidth}
        \begin{itemize}[leftmargin=*, nosep, topsep=0pt, itemsep=0pt, partopsep=0pt, after=\vspace{4pt}]
            \item Structured mealtime routines with visual schedules
            \item Meal tracking and progress dashboards for caregivers
            \item Customizable feeding plans for behavioral and sensory needs
            \item Motivational prompts, goal-setting and reward systems
        \end{itemize}
        \end{minipage} \\
        \hline
    \end{tabular}
    \label{tab_coreFeatures}
\end{table}

\begin{figure}[H]
    \centering
    \subfloat[\textbf{Gamified counting activity with food items}]{
        \includegraphics[width=0.45\textwidth]{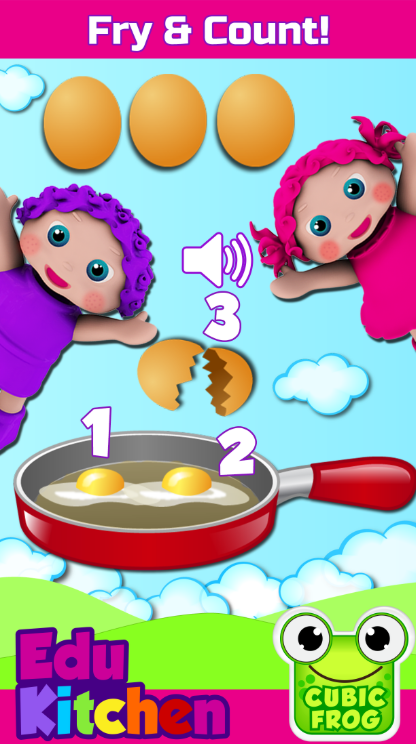}
        \label{fig:edukitchen_a}
    }
    \hspace{0.5em}
    \subfloat[\textbf{Interactive recycling task with food-related waste}]{
        \includegraphics[width=0.45\textwidth]{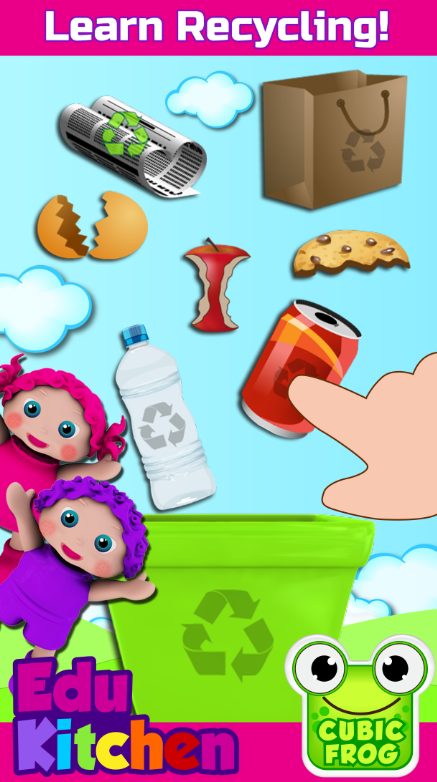}
        \label{fig:edukitchen_b}
    }  
    \caption{Screenshots from \textit{EduKitchen–Toddlers Food Games}. The app features food-themed games intended to build early familiarity with food through sorting, counting, and categorization.}
    \label{fig_edukitchen}
\end{figure}

\textit{EduKitchen–Toddlers Food Games} targets early childhood learners and emphasizes food familiarity through interactive play. The app includes mini-games such as frying and counting eggs (Figure \ref{fig:edukitchen_a}) and sorting food-related waste (Figure \ref{fig:edukitchen_b}). The tasks are designed to promote cognitive engagement with food concepts through bright visuals and immediate feedback. 

\begin{figure}[H]
    \centering
    \subfloat[\textbf{Prompt for emotional awareness during meals}]{
        \includegraphics[width=0.9\textwidth]{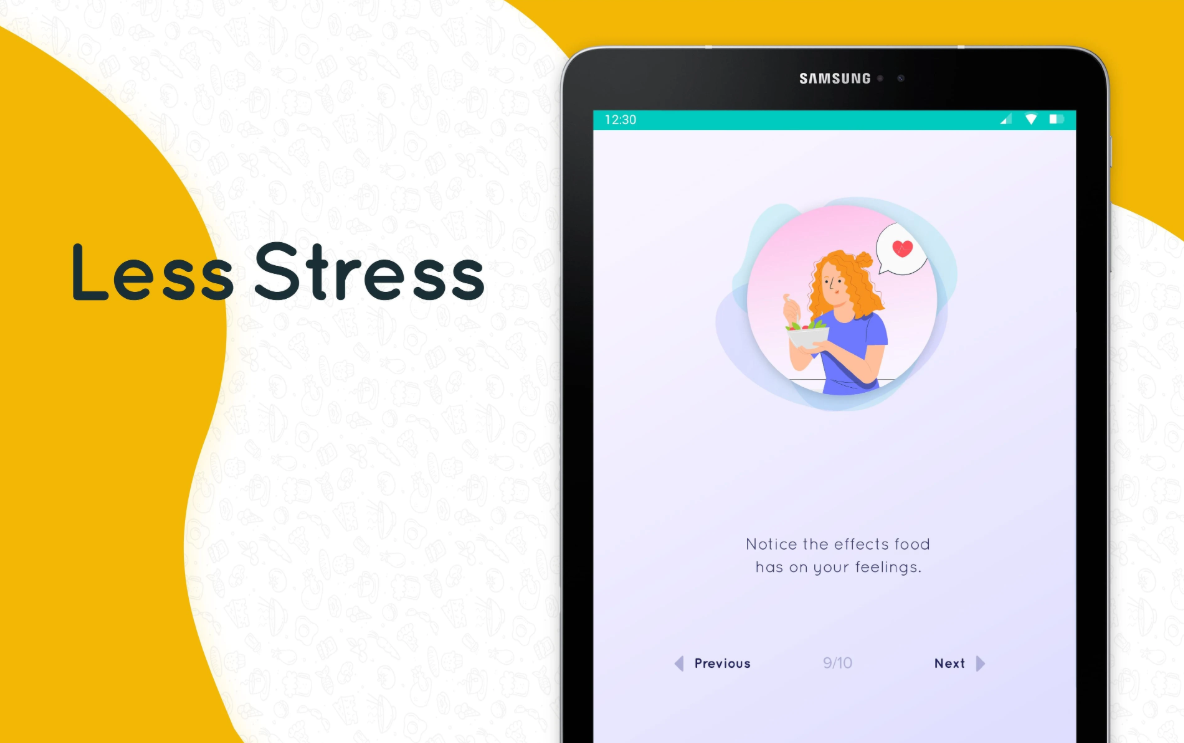}
        \label{fig:foodcoach2a}
    }\\[1em]
    \subfloat[\textbf{Suggestion to enhance the sensory environment with music}]{
        \includegraphics[width=0.9\textwidth]{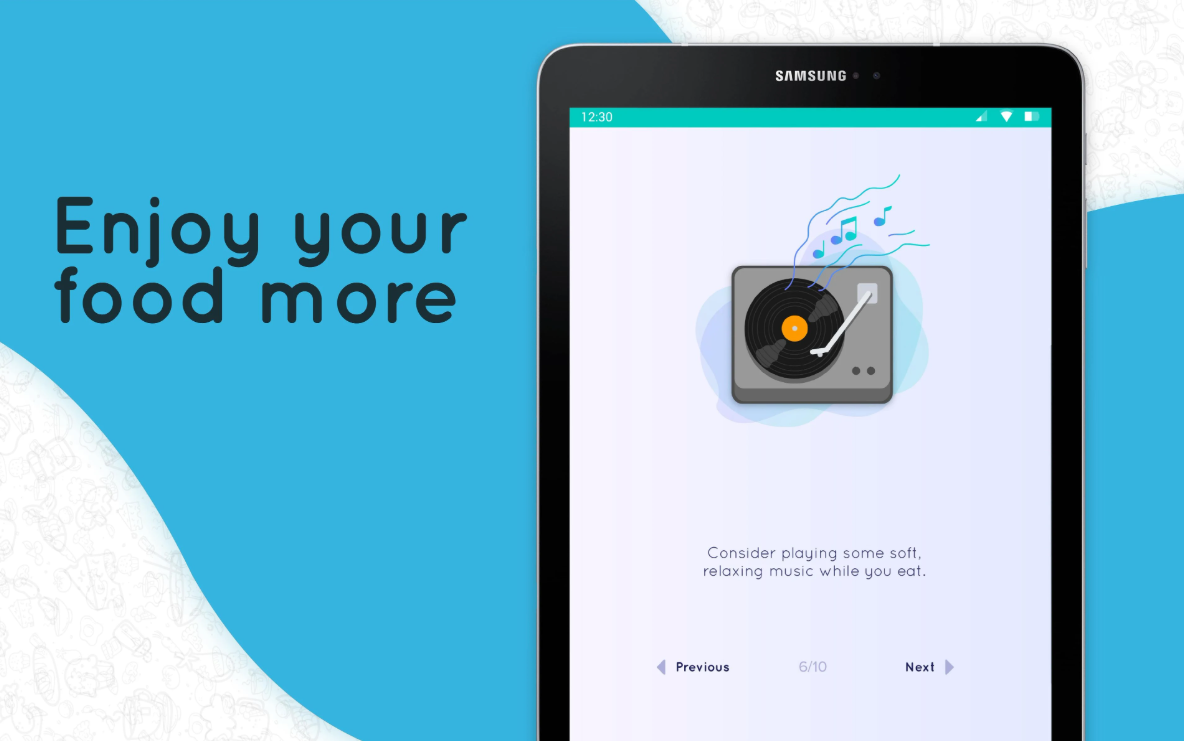}
        \label{fig:foodcoach2b}
    }
    \caption{Screenshots from \textit{Autism Food Coach 2}. The top image encourages users to reflect on how food affects their feelings, while the bottom image promotes calm sensory environments by suggesting soft background music during meals.}
    \label{fig_foodcoach}
\end{figure}

\textit{Autism Food Coach 2} is designed for caregivers and provides a structured approach to support feeding routines. As shown in Figure~\ref{fig_foodcoach}, the app includes motivational messages that encourage emotional reflection and sensory modulation during meals. For example, Figure \ref{fig:foodcoach2a} advises users to ``notice the effects food has on your feelings'', while Figure \ref{fig:foodcoach2b} suggests playing relaxing music during meals to enhance enjoyment. The app includes additional functionalities, such as progress tracking, visual schedules, and personalized meal planning tools.

\subsection{Feeding-Related Functionality}

Although both applications incorporate food-related content, they diverge considerably in functional purpose and clinical utility: one offers indirect exposure through child-led play, while the other provides direct support for caregiver-led interventions. This section evaluates the extent to which the apps support structured feeding interventions, promote adaptive mealtime behaviors, and guide caregiver decision-making.

\textit{EduKitchen–Toddlers Food Games} uses gamified tasks that promote food familiarity through child-led interaction. Tasks such as food sorting or basic cooking simulations can contribute to early exposure to food. However, they lack structured intervention strategies to address feeding challenges such as food neophobia (refusal), selectivity, or anxiety-related avoidance. The app's role in feeding intervention is thus indirect, emphasizing cognitive and sensory engagement rather than measurable behavioral change. The involvement of the caregiver is limited to adjusting the activity parameters, with no access to feedback systems or guidance tools.

In contrast, \textit{Autism Food Coach 2} integrates features explicitly designed to support caregivers in managing structured feeding routines. The features include customizable meal plans, progress dashboards, and visual schedules that allow systematic tracking of mealtime behaviors. Motivational prompts and sensory regulation suggestions (e.g., relaxing music or emotional reflection) are also used to create calmer mealtime environments. These evidence-informed functionalities enable data-driven caregiving by allowing parents to observe behavioral trends, adapt strategies, and systematically reinforce desired mealtime behaviors.

We formalize this analytical comparison in Textbox 1, which categorizes app features based on their ability to provide actionable guidance to support caregivers during feeding interventions. The analysis confirms that only \textit{Autism Food Coach 2} displays features that align with caregiver-mediated feeding interventions.

\begin{tcolorbox}[colback=white, colframe= black]
\begin{center}
\textbf{Textbox 1. App features categorized by caregiver decision-support capabilities}
\end{center}

\vspace{-0.5em} 
\noindent \textbf{Features with caregiver decision-support}
\begin{itemize}[leftmargin=*, nosep, topsep=0pt, itemsep=0pt, partopsep=0pt, after=\vspace{0.5em}]
  \item \textit{Instant feedback and rewards:} Reinforces positive child behaviors, allowing parents to adjust strategies in real time.
  \item \textit{Parent support tools:} Offers structured feedback, coaching prompts, and meal tracking to help manage feeding routines.
  \item \textit{Customizable progress monitoring:} Provides visual progress charts and behavior tracking to support informed parental decisions.
\end{itemize}

\noindent \textbf{Features without caregiver decision-support}
\begin{itemize}[leftmargin=*, nosep, topsep=0pt, itemsep=0pt, partopsep=0pt]
  \item \textit{Interactive child engagement:} Focuses on engaging the child through games and narratives but lacks direct caregiver guidance.
  \item \textit{Routine support lacking tracking systems:} Encourages routine-building but does not include tools for decision-making support.
\end{itemize}
\end{tcolorbox}

\subsection{Intervention Function Mapping Using the BCW Framework}

The Behavior Change Wheel (BCW) provides a systematic model for designing interventions based on behavioral determinants. Table \ref{tab_bcw_mapping} summarizes the BCW intervention functions identified in each application. We identified seven of the nine BCW functions across the two apps. It is worth noting that neither incorporated \textit{coercion} or \textit{restriction}.

\textit{EduKitchen–Toddlers Food Games} mapped onto five BCW functions: \textit{education}, \textit{training}, \textit{enablement}, \textit{incentivization}, and \textit{modeling}. Educational content is delivered through interactive lessons introducing food types and healthy eating. Training is embedded in mini-games that reinforce skills such as sorting and categorization. Enablement is supported via customizable gameplay settings to accommodate diverse sensory or developmental needs. Incentivization occurs through gamified rewards (e.g., praise animations), while modeling is reflected in the use of animated characters demonstrating appropriate feeding behaviors.

\textit{Autism Food Coach 2} incorporated six intervention functions: \textit{education}, \textit{training}, \textit{enablement}, \textit{persuasion}, \textit{incentivization}, and \textit{environmental restructuring}. The app provides educational content through guidance on balanced nutrition and mindful eating. Training is facilitated through structured visual schedules that model consistent feeding routines. Enablement includes personalized meal planning and caregiver-facing tracking tools. Persuasion is operationalized through motivational messages and positive affirmations. Incentivization is achieved through visual progress tracking and behavioral goal-setting. Environmental restructuring is implemented through tools that help caregivers create predictable low-stress mealtime environments (e.g., use of music, calming messages, and visual schedules).

Our mapping demonstrates that \textit{Autism Food Coach 2} uses a broader range of intervention functions, particularly those aimed directly at supporting caregiver behavior (\textit{persuasion, environmental restructuring}), which aligns with its focus on structured, caregiver-mediated intervention.

\begin{table}[t]
    \centering
    \caption{BCW intervention function mapping for the reviewed applications.}
    \footnotesize
    \begin{tabularx}{\textwidth}{|X|c|c|}
        \hline
        & \textit{EduKitchen–Toddlers Food Games} & \textit{Autism Food Coach 2} \\
        \hline
        \textbf{Education} & \cmark & \cmark \\ \hline
        \textbf{Training} & \cmark & \cmark \\ \hline
        \textbf{Enablement} & \cmark & \cmark \\ \hline
        \textbf{Persuasion} & \xmark & \cmark \\ \hline
        \textbf{Incentivization} & \cmark & \cmark \\ \hline
        \textbf{Modeling} & \cmark & \xmark \\ \hline
        \textbf{Environmental Restructuring} & \xmark & \cmark \\ \hline
        \textbf{Coercion} & \xmark & \xmark \\ \hline
        \textbf{Restriction} & \xmark & \xmark \\ \hline
    \end{tabularx}
    \label{tab_bcw_mapping}
\end{table}

\subsection{MARS Quality Assessment}
\begin{figure}[t]
    \centering
    \includegraphics[width=\textwidth]{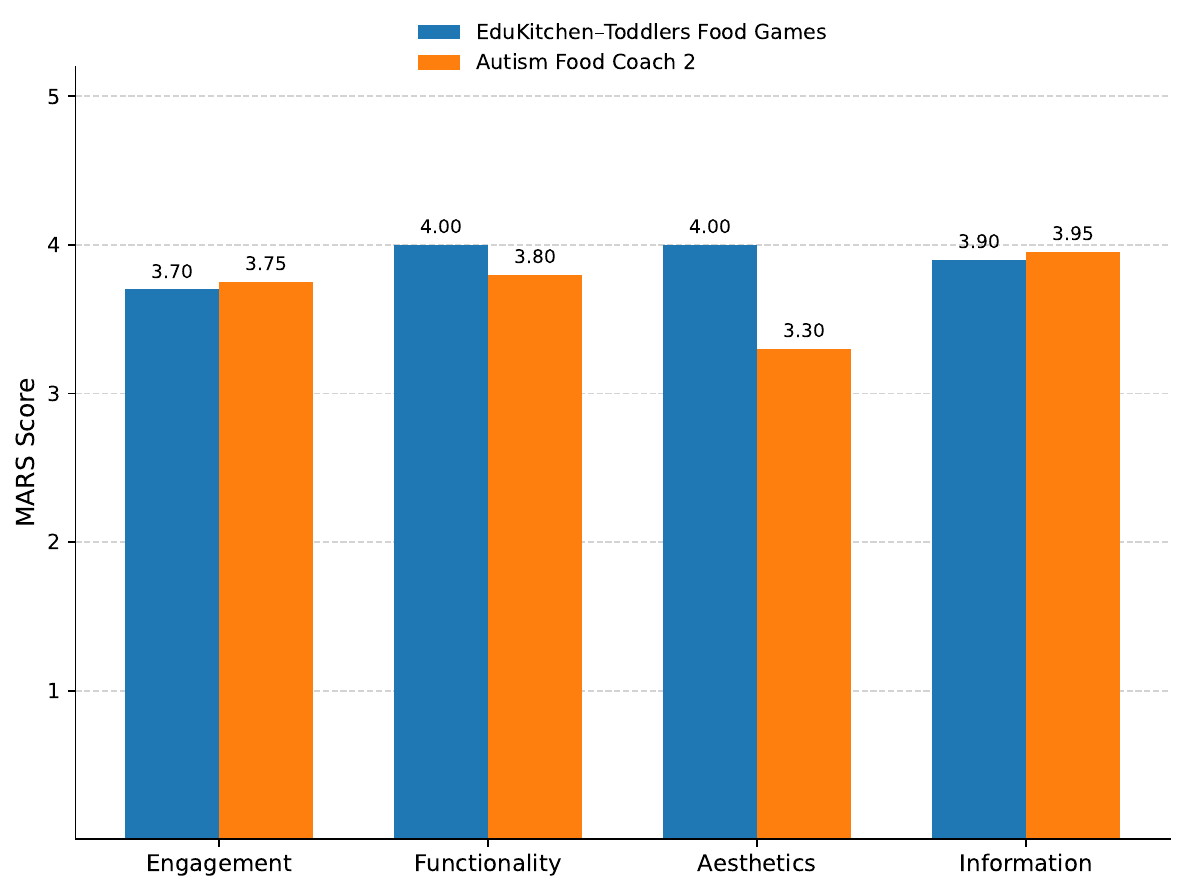}
    \caption{Comparison of Mobile App Rating Scale (MARS) quality dimension scores between \textit{EduKitchen–Toddlers Food Games} and \textit{Autism Food Coach 2}.}
    \label{fig:mars_quality_comparison}
\end{figure}

We present the results of the Mobile App Rating Scale (MARS) evaluation, based on consensus ratings provided by two independent reviewers who completed standardized MARS training. Table \ref{tab_app_ratings_mars_scale} summarizes the scores across the four MARS quality dimensions, i.e., \textit{engagement}, \textit{functionality}, \textit{aesthetics}, and \textit{information}, along with the overall mean and subjective quality ratings. Figure \ref{fig:mars_quality_comparison} provides a visual comparison of the two applications. The chart highlights a significant difference in the \textit{aesthetics} dimension, where \textit{EduKitchen–Toddlers Food Games} received a substantially higher score (4.0 compared to 3.3). In contrast, both applications performed comparably in \textit{engagement}, \textit{functionality}, and \textit{information}, with differences of 0.2 points or less.

\begin{table}[H]
    \centering
    \caption{MARS results for the two reviewed apps. Each dimension was scored on a 5-point scale.}
    \footnotesize
    \begin{tabularx}{\textwidth}{|X|c|c|}
        \hline
        & \textit{EduKitchen–Toddlers Food Games} & \textit{Autism Food Coach 2} \\
        \hline
        \textbf{Platform} & iOS & iOS \\ \hline
        \textbf{Engagement} & 3.7 & 3.75 \\ \hline
        \textbf{Functionality} & 4.0 & 3.8 \\ \hline
        \textbf{Aesthetics} & 4.0 & 3.3 \\ \hline
        \textbf{Information} & 3.9 & 3.95 \\ \hline
        \textbf{Overall Score} & 3.9 & 3.7 \\ \hline
        \textbf{Subjective Quality Score} & 4.0 & 4.0 \\ \hline
    \end{tabularx}
    \label{tab_app_ratings_mars_scale}
\end{table}

\textit{EduKitchen–Toddlers Food Games} achieved an average MARS score of 3.9. It performed strongest in \textit{functionality} (4.0) and \textit{aesthetics} (4.0), which reflects a polished and responsive interface with visually engaging content. \textit{Engagement} (3.7) and \textit{information quality} (3.9) also received strong ratings, suggesting an effective interaction design and age-appropriate content. The subjective quality score of 4.0 further implies a favorable overall impression.

\textit{Autism Food Coach 2} received a slightly lower average score of 3.7. Its highest ratings included \textit{information} (3.95) and \textit{engagement} (3.75), driven by relevant content and caregiver-centered functionality. \textit{Functionality} scored 3.8, while \textit{aesthetics} received the lowest rating (3.3), attributed to its more utilitarian visual design. Despite this, the subjective quality score remained high at 4.0, underscoring the perceived usefulness of its structured tools.

The MARS evaluation highlighted key strengths and limitations with respect to both applications. \textit{EduKitchen} excelled in visual engagement and interactivity, making it appealing for children. However, its lack of clinical validation and limited caregiver customization reduce its potential for therapeutic application. In contrast, \textit{Autism Food Coach 2} provided evidence-informed features that support caregiver-mediated feeding routines but would benefit from improved visual design and rigorous clinical validation. Table \ref{tab_structured} summarizes the strengths and limitations of both apps.

\begin{table}[H]
    \centering
    \caption{Strengths and limitations of the two reviewed apps.}
    \footnotesize
    \begin{tabularx}{\textwidth}{|l|X|X|}
        \hline
        & \textit{EduKitchen–Toddlers Food Games} & \textit{Autism Food Coach 2} \\
        \hline
        \textbf{Strengths} & Provides an engaging and interactive learning experience with high-quality graphics and food-themed mini-games; includes basic caregiver customization. & Offers tailored meal plans, progress tracking, and structured coaching tools that support caregiver-led feeding interventions. \\
        \hline
        \textbf{Limitations} & No scientific validation; limited features for caregiver-led interventions or clinical feeding challenges. & Limited visual design quality; lacks empirical validation through clinical trials. \\
        \hline
    \end{tabularx}
    \label{tab_structured}
\end{table}
\section{Discussion}
\label{discussion}
This systematic review examined the Apple App Store and Google Play Store to identify and evaluate mobile applications addressing feeding challenges in children with autism spectrum disorder (ASD). Out of over 300 screened applications, only two (\textit{EduKitchen–Toddlers Food Games} and \textit{Autism Food Coach 2}) met the inclusion criteria. The scarcity of qualifying applications reveals a substantial gap in available mHealth apps specifically designed to address feeding difficulties among children with ASD.

The two applications differ in target users and intervention logic. \textit{EduKitchen–Toddlers Food Games} adopts a child-centered approach, using gamified tasks to increase familiarity with food and promote positive mealtime engagement. In contrast, \textit{Autism Food Coach 2} is caregiver-oriented and emphasizes structured routine support through planning and monitoring tools. Mapping these applications to the BCW framework revealed that \textit{EduKitchen–Toddlers Food Games} aligned with five intervention functions (\textit{education}, \textit{training}, \textit{enablement}, \textit{incentivization}, and \textit{modeling}), which reflects its emphasis on skill-building through interactive play. whereas \textit{Autism Food Coach 2} aligned with six: \textit{education}, \textit{training}, \textit{enablement}, \textit{persuasion}, \textit{incentivization}, and \textit{environmental restructuring}. Our findings reflect the distinct target audiences and use cases of each application and highlight the utility of evaluating the intervention logic of mobile health interventions using frameworks such as the BCW \cite{michie_behaviour_2011}. The quality assessment conducted using the MARS framework \cite{stoyanov2015mobile} revealed that both applications demonstrated moderate to high quality. \textit{EduKitchen–Toddlers Food Games} received a marginally higher overall score (3.9) than \textit{Autism Food Coach 2} (3.7), with particular strengths in functionality and visual appeal. In contrast, \textit{Autism Food Coach 2} offered a greater alignment with evidence-based feeding practices. Notably, neither app provided scientific validation or included robust customization features to adapt to individual child needs and profiles, an important limitation given the heterogeneity of feeding issues in children with ASD.

Compared to nutrition apps evaluated using MARS (mean score = 2.95) \cite{Diet-Related_Mobile_Apps}, both autism-focused apps scored higher in quality but showed a narrower intervention scope. Whereas nutrition apps often emphasize dietary tracking, social engagement, and personalized feedback, the autism apps focused primarily on gamified exposure and routine support. This limited scope may constrain their effectiveness in addressing the multifactorial drivers of feeding difficulties in ASD, including sensory sensitivities, oral-motor challenges, and behavioral rigidity \cite{larco_review_2018}. While educational content and routine-building tools are valuable, future app development should incorporate modules that explicitly support sensory integration, oral-motor skill acquisition, and graduated exposure to novel foods.

While this review provides a structured assessment of the current landscape, several limitations must be considered when interpreting these findings. The inclusion threshold of a minimum of 100 downloads may have excluded novel or niche applications with limited visibility. In addition, app availability was influenced by the geographic location of the reviewers' devices, which could have led to omitting region-specific apps not accessible in the Africa region. Although extensive search terms were used, some relevant apps may have been missed due to alternative labeling by developers. Furthermore, restricted features behind paywalls could not be fully evaluated. Lastly, while MARS provides a standardized assessment from a rater perspective, it may not fully capture caregiver preferences and usability. Future work could consider user-centered measures, such as uMARS, to assess app acceptability and perceived usefulness.

Future research should broaden the search strategy to mitigate geographic sampling bias, for example, by using multiregion app store accounts to more comprehensively capture global app availability. Co-design approaches that involve parents, clinicians and speech or occupational therapists will be crucial in ensuring that future applications address both clinical needs and contextual realities. Particularly in resource-constrained settings, culturally tailored and clinically validated mHealth apps are urgently needed to support caregivers to manage ASD-related feeding challenges. Longitudinal studies evaluating behavioral outcomes and caregiver burden can further establish the real-world impact of these digital interventions.

\section{Conclusion}
\label{conclusion}

This study highlights a critical shortage of mobile health (mHealth) applications specifically designed to support feeding challenges in children with autism spectrum disorder (ASD). Using the Behavior Change Wheel (BCW) and Mobile App Rating Scale (MARS) frameworks, we evaluated the functional strengths and theoretical underpinnings of the two eligible apps. Although both applications incorporated behavior change techniques and demonstrated moderate to high usability, neither was grounded in clinically validated feeding interventions. These findings underscore the need for user-centered, evidence-based app development that integrates behavioral, sensory, and skill-building supports tailored to the diverse needs of children with ASD. Expanding access to effective, clinically validated, and culturally responsive mHealth tools is essential for advancing digital health equity, particularly in low-resource settings where traditional intervention services may be limited.

\section*{Summary Table}
\begin{itemize}
    \item This systematic review identified a critical gap in the mHealth landscape: a severe scarcity of apps specifically designed for feeding difficulties in autism, with only two of over 300 screened apps meeting inclusion criteria.
    \item The evaluated apps, while demonstrating moderate-to-high usability (MARS scores: 3.7-3.9), revealed a narrow focus. Their design, mapped via the Behavior Change Wheel, prioritized education and routine support but overlooked core autism-specific challenges like sensory sensitivities.
    \item A critical limitation is the lack of scientific validation and robust customization, which highlights a significant disconnect between app development and evidence-based autism feeding interventions.
    \item Future efforts should prioritize co-design with caregivers and clinicians to create clinically grounded, comprehensive, and accessible mHealth apps, especially for low-resource settings.
\end{itemize}

\section*{Declaration of Competing Interests}
The authors declare that they have no known competing financial interests or personal relationships that could have appeared to influence the work reported in this paper.

\section*{Acknowledgements}
This research was supported by the Consortium for Advanced Research Training in Africa (CARTA). CARTA is jointly led by the African Population and Health Research Center and the University of the Witwatersrand and funded by the Carnegie Corporation of New York (Grant No. G-21-58722 and G-PS-23-60922), Sida (Grant No: 16604), Norwegian Agency for Development Cooperation (Norad) (Grant No: QZA-21/0162), Oak Foundation (Grant No. OFIL-24-091) and the Science for Africa Foundation to the Developing Excellence in Leadership, Training and Science in Africa (DELTAS Africa) programme (Del-22-006) with support from Wellcome and the UK Foreign, Commonwealth \& Development Office and is part of the EDCPT2 programme supported by the European Union. The statements made and views expressed are solely the responsibility of the Author.

\bibliographystyle{IEEEtran} 
\bibliography{main}

\begin{thebibliography}{10}
\providecommand{\url}[1]{#1}
\csname url@samestyle\endcsname
\providecommand{\newblock}{\relax}
\providecommand{\bibinfo}[2]{#2}
\providecommand{\BIBentrySTDinterwordspacing}{\spaceskip=0pt\relax}
\providecommand{\BIBentryALTinterwordstretchfactor}{4}
\providecommand{\BIBentryALTinterwordspacing}{\spaceskip=\fontdimen2\font plus
\BIBentryALTinterwordstretchfactor\fontdimen3\font minus \fontdimen4\font\relax}
\providecommand{\BIBforeignlanguage}[2]{{%
\expandafter\ifx\csname l@#1\endcsname\relax
\typeout{** WARNING: IEEEtran.bst: No hyphenation pattern has been}%
\typeout{** loaded for the language `#1'. Using the pattern for}%
\typeout{** the default language instead.}%
\else
\language=\csname l@#1\endcsname
\fi
#2}}
\providecommand{\BIBdecl}{\relax}
\BIBdecl

\bibitem{cdc_autism_2023}
\BIBentryALTinterwordspacing
{CDC}. {Autism Spectrum Disorder (ASD)}. [Online]. Available: \url{https://www.cdc.gov/autism/data-research/}
\BIBentrySTDinterwordspacing

\bibitem{who_autism_2023}
\BIBentryALTinterwordspacing
{WHO}. Autism. [Online]. Available: \url{https://www.who.int/news-room/fact-sheets/detail/autism-spectrum-disorders}
\BIBentrySTDinterwordspacing

\bibitem{nadon_feeding_2013}
\BIBentryALTinterwordspacing
G.~Nadon, D.~Feldman, and E.~Gisel, ``Feeding issues associated with the autism spectrum disorders,'' in \emph{Recent Advances in Autism Spectrum Disorders}, M.~Fitzgerald, Ed.\hskip 1em plus 0.5em minus 0.4em\relax Rijeka: IntechOpen, 2013, ch.~25. [Online]. Available: \url{https://doi.org/10.5772/53644}
\BIBentrySTDinterwordspacing

\bibitem{marshall_features_2014}
J.~Marshall, R.~J. Hill, J.~Ziviani, and P.~Dodrill, ``Features of feeding difficulty in children with autism spectrum disorder,'' \emph{International journal of speech-language pathology}, vol.~16, no.~2, pp. 151--158, 2014.

\bibitem{leader_feeding_2020}
G.~Leader, E.~Tuohy, J.~L. Chen, A.~Mannion, and S.~P. Gilroy, ``Feeding problems, gastrointestinal symptoms, challenging behavior and sensory issues in children and adolescents with autism spectrum disorder,'' \emph{Journal of autism and developmental disorders}, vol.~50, pp. 1401--1410, 2020.

\bibitem{xiao_correlation_2016}
\BIBentryALTinterwordspacing
X.~Liu, J.~Liu, X.~Xiong, T.~Yang, N.~Hou, X.~Liang, J.~Chen, Q.~Cheng, and T.~Li, ``Correlation between nutrition and symptoms: Nutritional survey of children with autism spectrum disorder in chongqing, china,'' \emph{Nutrients}, vol.~8, no.~5, 2016. [Online]. Available: \url{https://www.mdpi.com/2072-6643/8/5/294}
\BIBentrySTDinterwordspacing

\bibitem{marinov_micronutrient_2024}
\BIBentryALTinterwordspacing
D.~Marinov, R.~Chamova, and R.~Pancheva, ``Micronutrient deficiencies in children with autism spectrum disorders compared to typically developing children – a scoping review,'' \emph{Research in Autism Spectrum Disorders}, vol. 114, p. 102396, 2024. [Online]. Available: \url{https://www.sciencedirect.com/science/article/pii/S1750946724000710}
\BIBentrySTDinterwordspacing

\bibitem{sharp_feeding_2013}
W.~G. Sharp, R.~C. Berry, C.~McCracken, N.~N. Nuhu, E.~Marvel, C.~A. Saulnier, A.~Klin, W.~Jones, and D.~L. Jaquess, ``Feeding problems and nutrient intake in children with autism spectrum disorders: a meta-analysis and comprehensive review of the literature,'' \emph{Journal of autism and developmental disorders}, vol.~43, pp. 2159--2173, 2013.

\bibitem{Rojas-Torres2020}
\BIBentryALTinterwordspacing
L.~Rojas-Torres, Y.~Alonso-EstebanORCID, and F.~Alcantud-Marín, ``Early intervention with parents of children with autism spectrum disorders: A review of programs,'' \emph{Children}, 2020. [Online]. Available: \url{https://doi.org/10.3390/children7120294}
\BIBentrySTDinterwordspacing

\bibitem{Beaudoin2014}
\BIBentryALTinterwordspacing
A.~Beaudoin, G.~Sébire, and M.~Couture, ``Parent training interventions for toddlers with autism spectrum disorder,'' \emph{Autism Research and Treatment}, vol. 2014, 2014. [Online]. Available: \url{https://doi.org/10.1155/2014/839890}
\BIBentrySTDinterwordspacing

\bibitem{bharat_mhealth_2022}
R.~Bharat, U.~Uzaina, T.~Yadav, S.~Niranjan, and P.~Kurade, ``mhealth apps delivering early intervention to support parents of children with autism: a scoping review protocol,'' \emph{BMJ paediatrics open}, vol.~6, no.~1, p. e001358, 2022.

\bibitem{Dunn2017}
\BIBentryALTinterwordspacing
R.~Dunn, J.~Elgart, L.~Lokshina, A.~Faisman, E.~Khokhlovich, Y.~Gankin, and A.~Vyshedskiy, ``Performance of children with autism in parent-administered cognitive and language exercises,'' \emph{bioRxiv}, 2017. [Online]. Available: \url{https://doi.org/10.1101/146449}
\BIBentrySTDinterwordspacing

\bibitem{aziz_study_2019}
N.~S.~A. Aziz, W.~F.~W. Ahmad, and A.~S. Hashim, ``A study on mobile applications developed for children with autism,'' in \emph{Recent Trends in Data Science and Soft Computing}, F.~Saeed, N.~Gazem, F.~Mohammed, and A.~Busalim, Eds.\hskip 1em plus 0.5em minus 0.4em\relax Cham: Springer International Publishing, 2019, pp. 772--780.

\bibitem{law_the_2018}
\BIBentryALTinterwordspacing
G.~C. Law, M.~Neihart, and A.~Dutt, ``The use of behavior modeling training in a mobile app parent training program to improve functional communication of young children with autism spectrum disorder,'' \emph{Autism}, vol.~22, no.~4, pp. 424--439, 2018, pMID: 29153002. [Online]. Available: \url{https://doi.org/10.1177/1362361316683887}
\BIBentrySTDinterwordspacing

\bibitem{bonnot_mobile_2021}
O.~Bonnot, V.~Adrien, V.~Venelle, D.~Bonneau, F.~Gollier-Briant, and S.~Mouchabac, ``Mobile app for parental empowerment for caregivers of children with autism spectrum disorders: Prospective open trial,'' \emph{JMIR Ment Health}, vol.~8, no.~9, p. e27803, Sep 2021.

\bibitem{larco_review_2018}
A.~Larco, F.~Enríquez, and S.~Luján-Mora, ``Review and evaluation of special education ios apps using mars,'' in \emph{2018 IEEE World Engineering Education Conference (EDUNINE)}, 2018, pp. 1--6.

\bibitem{Bradshaw2022}
\BIBentryALTinterwordspacing
J.~Bradshaw, K.~Wolfe, R.~Hock, and L.~Scopano, ``Advances in supporting parents in interventions for autism spectrum disorder,'' \emph{Pediatric Clinics of North America}, 2022. [Online]. Available: \url{https://doi.org/10.1016/j.pcl.2022.04.002}
\BIBentrySTDinterwordspacing

\bibitem{Houser2014}
\BIBentryALTinterwordspacing
L.~Houser, M.~McCarthy, L.~Lawer, and D.~Mandell, ``A challenging fit: Employment, childcare, and therapeutic support in families of children with autism spectrum disorders,'' \emph{Journal of Social Service Research}, 2014. [Online]. Available: \url{https://doi.org/10.1080/01488376.2014.930944}
\BIBentrySTDinterwordspacing

\bibitem{michie_behaviour_2011}
S.~Michie, M.~M. Van~Stralen, and R.~West, ``The behaviour change wheel: a new method for characterising and designing behaviour change interventions,'' \emph{Implementation science}, vol.~6, pp. 1--12, 2011.

\bibitem{stoyanov2015mobile}
S.~R. Stoyanov, L.~Hides, D.~J. Kavanagh, O.~Zelenko, D.~Tjondronegoro, and M.~Mani, ``Mobile app rating scale: a new tool for assessing the quality of health mobile apps,'' \emph{JMIR mHealth and uHealth}, vol.~3, no.~1, p. e3422, 2015.

\bibitem{ethnologue_english_2024}
\BIBentryALTinterwordspacing
D.~Eberhard, G.~Simons, and C.~Fennig, ``Ethnologue: Languages of the world, 27th edition,'' 2024. [Online]. Available: \url{https://www.ethnologue.com/language/eng}
\BIBentrySTDinterwordspacing

\bibitem{huckvale2015smartphone}
\BIBentryALTinterwordspacing
K.~Huckvale, S.~Adomaviciute, J.~Prieto, M.~Leow, and J.~Car, ``Smartphone apps for calculating insulin dose: a systematic assessment,'' \emph{BMC Medicine}, 2015. [Online]. Available: \url{https://bmcmedicine.biomedcentral.com/articles/10.1186/s12916-015-0314-7}
\BIBentrySTDinterwordspacing

\bibitem{liberati_prisma_2009}
\BIBentryALTinterwordspacing
A.~Liberati, D.~G. Altman, J.~Tetzlaff, C.~Mulrow, P.~C. Gøtzsche, J.~P.~A. Ioannidis, M.~Clarke, P.~J. Devereaux, J.~Kleijnen, and D.~Moher, ``The prisma statement for reporting systematic reviews and meta-analyses of studies that evaluate health care interventions: Explanation and elaboration,'' \emph{PLOS Medicine}, vol.~6, no.~7, pp. 1--28, 07 2009. [Online]. Available: \url{https://doi.org/10.1371/journal.pmed.1000100}
\BIBentrySTDinterwordspacing

\bibitem{Diet-Related_Mobile_Apps}
\BIBentryALTinterwordspacing
J.~Choi, C.~Chung, and H.~Woo, ``Diet-related mobile apps to promote healthy eating and proper nutrition: A content analysis and quality assessment,'' \emph{International Journal of Environmental Research and Public Health}, 2021. [Online]. Available: \url{https://www.mdpi.com/1660-4601/18/7/3496}
\BIBentrySTDinterwordspacing

\end{thebibliography}

\clearpage
\onecolumn
\appendix
\renewcommand{\thetable}{A.\arabic{table}}
\setcounter{table}{0}
\section*{Appendix A. List of Screened Apps}
\label{appendixScreenedApps}

\setlength{\LTleft}{0pt}
\setlength{\LTright}{0pt}

\begin{longtable}{|p{0.8cm}|p{4cm}|p{1.8cm}|p{6cm}|}
\caption{Screened applications and their primary focus.} \label{tab_screened_apps} \\

\hline
\textbf{} & \textbf{Apps Name} & \textbf{Platform} & \textbf{Primary Focus} \\
\hline
\endfirsthead

\multicolumn{4}{c}%
{\tablename\ \thetable\ (\textit{Continued}) } \\
\hline
\textbf{} & \textbf{Apps Name} & \textbf{Platform} & \textbf{Primary Focus} \\
\hline
\endhead

\hline \multicolumn{4}{r}{\textit{Continued on next page}} \\
\endfoot

\hline
\endlastfoot

1 & 2nd Grade Baby Book Animal Flashcards For Kids or Kindergarten to Learn Words With Sounds & iOS & To help young children learn first words and animal names through interactive flashcards with sounds \\ \hline
2 & Aiko and Egor: Animation 4 Autism & Android & To promote learning through animated videos and games targeting key skills for individuals with autism \\ \hline
3 & App4Autism-Timer, Visual Planning & Android & To help parents and teachers efficiently use tools like images and audios to work with children with autism \\ \hline
4 & ASDHelp: Kids's Autism Games & Android & To help therapists track child's progress and suggest new exercises, and help parents reduce the effects of autism at home \\ \hline
5 & ASDetect & Android & To help parents and caregivers review possible early signs of autism in children under 2.5 years \\ \hline
6 & Autism 360 & Android & The parenting app for parenting kids 1-18 in autism spectrum and associated conditions \\ \hline
7 & Autism 911 & Android & To support families with autistic children \\ \hline
8 & Autism ABC App & Android & To provide a dedicated tool for educators and parents of autistic children, offering both entertainment and therapeutic support \\ \hline
9 & Autism AI & Android & An autistic traits detection system using AI \\ \hline
10 & Autism Diary & Android & To track daily habits and behaviors, helping caregivers manage the needs of individuals with autism \\ \hline
11 & Autism Evaluation Checklist & Android & To help in testing for autistic spectrum disorder \\ \hline
12 & Autism Food Coach 2 & iOS & To support parents in helping children develop healthier eating habits through mindful eating practices \\ \hline
13 & Autism Parenting Magazine & Android & To provide autism therapies, solutions, and news \\ \hline
14 & Autism Social Video Exercises & Android & Designed to build social confidence for individuals with ASD through interactive video exercises \\ \hline
15 & Autism Speaks Fundraising & Android & To help in fundraising by tracking progress and personalizing fundraising pages for autistic people \\ \hline
16 & Autism Speech and Language & Android & For parents to help their autistic children improve speech and language development \\ \hline
17 & Autism Test (Adult) & Android & To evaluate the risk of autism spectrum disorder with a validated screening test \\ \hline
18 & Autism Test (Child) & Android & To test if a child is autistic \\ \hline
19 & AutiSpark: Kids Autism Games & Android & To provide fun learning activities and games specially for kids with autism spectrum disorder \\ \hline
20 & Autistapp & Android & App to help autistic individuals \\ \hline
21 & Autism: Daily Living and Caring & Android & Designed for families and carers to create a private group to help care for someone with autism \\ \hline
22 & Autimo-AMIKEO APPS & Android & To help children with autism understand emotions with enjoyable activities \\ \hline
23 & Awetism Insights & Android & App designed to support parents of autistic children \\ \hline
24 & Car Puzzle 2 for toddlers & iOS & To help toddlers, including those with autism, learn about cars and transportation through interactive puzzles \\ \hline
25 & Easy jigsaw puzzle games for toddlers and babies & iOS & Made for kids aged 1-4 to introduce food and kitchen items through puzzles and mini-games \\ \hline
26 & EduKitchen-Toddlers Food Games & iOS & To teach children aged 2-6 healthy eating, counting, and logic through kitchen-themed games \\ \hline
27 & Flashcards for kids-First Food Words & iOS & To provide customizable, interactive flashcards for early childhood learning (ages 0-6) \\ \hline
28 & Fruits and Vegetables flashcards quiz and matching game for toddlers and kids in English & iOS & To help kids learn names of fruits and vegetables through interactive games \\ \hline
29 & Fun Routine-Autism & Android & To manage daily routines, improve communication, and strengthen socio-emotional development \\ \hline
30 & iDo Food-Kids with special needs learn dining skills (Full Version) & iOS & To teach cooking/dining skills for individuals with autism and intellectual disabilities \\ \hline
31 & Jade & Android & For children/adolescents with autism, developmental delay, or learning difficulties \\ \hline
32 & Language Therapy for Children & Android & To provide early intervention language therapy for children with delays and autism \\ \hline
33 & Leeloo AAC-Autism Speech App & Android & To help non-verbal children communicate using visual cards (AAC/PECS) \\ \hline
34 & Learn Autism & Android & Free, on-demand videos created by Autism experts \\ \hline
35 & Matraquinha: Autism & Android & To help autistic people with communication \\ \hline
36 & Meet the Colors & iOS & To help toddlers learn colors through interactive games \\ \hline
37 & Meet the Letters Lowercase & iOS & To help toddlers learn lowercase letters through matching games \\ \hline
38 & Meet the Letter Uppercase & iOS & To help toddlers learn uppercase letters through interactive play \\ \hline
39 & Meet the Numbers & iOS & To help toddlers learn numbers 0-10 through tapping games \\ \hline
40 & Meet the Shapes & iOS & To teach toddlers about shapes using interactive games \\ \hline
41 & Music Therapy for Autism & Android & To help children with ASD and ADHD \\ \hline
42 & My Autism Navigator & Android & To support autistic children's learning in everyday activities \\ \hline
43 & NeuroSchemas for Autism & Android & To help individuals learn social rules for different situations \\ \hline
44 & Otsimo & Android & Assistive technology for special education \\ \hline
45 & Otsimo-Special Education & Android & Designed for individuals with autism, Down syndrome, and other special needs \\ \hline
46 & Otsimo-Speech Therapy SLP & Android & To help kids with speech issues, including autism \\ \hline
47 & Otsimo AAC-Tap and Talk & Android & AAC speech solution for individuals with communication difficulties \\ \hline
48 & Pediatric Nutrition & Android & Evidence-based guidance on childhood nutrition issues \\ \hline
49 & POOW The Food Hero & iOS & To assist parents of picky eaters (ages 2-5) to try new flavors \\ \hline
50 & Pravah: Autism Care Activities & Android & Personalized therapy activities for Autism/ADHD \\ \hline
51 & SenLife-ADHD and Autism Support & Android & To help parents track behavior of children with ADHD or ASD and create reports for home or school care \\ \hline
52 & Sight Words 1 Flashcards & iOS & To teach toddlers sight words through interactive flashcards \\ \hline
53 & Sight Words 1 Guessing Game & iOS & To help toddlers learn 16 sight words through flashcards \\ \hline
54 & Sight Words 2 Flashcards & iOS & To teach toddlers sight words using interactive flashcards \\ \hline
55 & Sight Words 3 Flashcards Sight Words 2 Guessing Game & iOS & To help toddlers learn sight words through interactive games \\ \hline
56 & Sight Words 3 Guessing Game & iOS & To teach sight words with tappable characters and animations \\ \hline
57 & TalkTablet PRO Autism and Stroke & Android & Comprehensive AAC speech solution for communication difficulties \\ \hline
58 & The Autism Helper & Android & To support educators and parents of autistic individuals \\ \hline
59 & ThinkAutism & Android & To learn about Asperger's and complete a screening questionnaire \\ \hline
60 & Tiny Tastes & iOS & To encourage picky eaters to try new foods through games \\ \hline
61 & Verbal Autism & Android & Designed for nonverbal children on the Autism Spectrum \\ \hline
\end{longtable}
\twocolumn 

\end{document}